\documentclass[aps,prl,twocolumn,showpacs]{revtex4}  
\usepackage{epsfig}  
 \begin{document}  
  
\title{Phonon-induced decay of the electron spin in quantum dots}  
\author{Vitaly N. Golovach, Alexander Khaetskii, and Daniel Loss}  
\address{Department of Physics and Astronomy, University of Basel,\\  
 Klingelbergstrasse 82, CH-4056 Basel, Switzerland}  
\newcommand{\bra}[1]{{\langle #1 |}}  
\newcommand{\ket}[1]{{| #1 \rangle}}  
\newcommand{\heading}[1]{%
 \begin{center}  
   \large\bf  \shadowbox{#1}%
 \end{center}  
 \vspace{1ex minus 1ex}}  
\newcommand{\BF}[1]{{\bf #1:}\hspace{1em}\ignorespaces}  

\date{\today}
\draft{}
\preprint{}
  
\begin{abstract}  
We study spin relaxation and decoherence in a   
 GaAs quantum dot due to spin-orbit interaction.  
We derive an effective Hamiltonian which couples   
 the electron spin to phonons or any other   
 fluctuation of the dot potential.  
We show that the spin decoherence time $T_2$ is as large   
 as the spin relaxation time $T_1$, under realistic conditions.  
For the Dresselhaus and Rashba spin-orbit couplings,   
 we find that, in leading order, the effective magnetic field can have 
 only fluctuations transverse to the applied magnetic field. 
As a result, $T_2=2T_1$ for arbitrarily   
 large Zeeman splittings, in contrast to the naively expected case 
 $T_2\ll T_1$.  
We show that the spin decay is drastically suppressed   
 for certain magnetic field directions and values of the  
 Rashba coupling constant.   
Finally, for the spin coupling to acoustic phonons, we show that  
 $T_2=2T_1$ for all spin-orbit mechanisms in leading  
 order in the electron-phonon interaction.  
\end{abstract}  
\maketitle  
  
Phase coherence of spin in quantum dots (QDs) is of central importance  
 for spin-based quantum computation in the solid 
 state~\cite{Loss97,Spintronics}.   
Sufficiently long coherence times are needed for implementing 
 quantum algorithms and error correction schemes.  
If the qubit is operated as a  classical  bit, its decay time is given  
 by the spin relaxation time $T_1$, which is the time of a spin-flip process.  
For quantum computation, however, the spin decoherence  
 time $T_2$ --- the lifetime of a coherent superposition of  spin up and  
 spin down states ---  must be sufficiently long.  
In semiconductor QDs, the spin coherence is limited by  
 the dot {\em intrinsic} degrees of freedom, such as   
 phonons, spins of nuclei,   
 excitations on the Fermi surface (e.g. in metallic gates),  
 fluctuating impurity states nearby the dot, electromagnetic fields, etc.  
It is well known (and experimentally verified) that the $T_1$ time of spin  
 in QDs is extremely long, extending up to $100\,\mu{\rm s}$.  
The decoherence time $T_2$, in its turn, is limited by both spin-flip   
 and dephasing processes, and can be much smaller than $T_1$, although   
 its upper bound is $T_2\leq 2T_1$.  
Knowledge of the mechanisms of spin relaxation and decoherence in QDs  
 can allow one to find regimes with the  least spin decay.  
  
Recently, the spin relaxation time $T_1$   
 in a single-electron GaAs QD  
 was measured~\cite{Hanson} by means of a pulsed relaxation measurement   
 technique (PRMT)~\cite{Fujisawa}.  
This technique was previously applied to measure the  
 triplet-to-singlet spin relaxation in a two-electron quantum   
 dot~\cite{Fujisawa1}, yielding a spin relaxation time of $200\,\mu{\rm s}$.  
The application of the PRMT to Zeeman sublevels became possible with  
 resolving the Zeeman splitting in {\em dc} transport   
 spectroscopy~\cite{Hanson,Potok}, which required a magnetic field   
 $B>5\,{\rm T}$.  
The results of Ref.~\onlinecite{Hanson} show that   
 $T_1>50\,\mu{\rm s}$ at   
 $B=7.5\,{\rm T}$ and $14\,{\rm T}$, with no indication of  
 a $B$-field dependence.  
Experimental values for the spin $T_2$ time in a single QD  
 are not available yet, but an ESR scheme for its measurement has been  
 proposed~\cite{EngelLossPRLPRB}.  
The ensemble spin decoherence time $T_2^*$ was measured in   
 $n$-doped GaAs bulk semiconductors~\cite{Kikkawa}, demonstrating  
 coherent spin precession over times exceeding $T_2^*\sim 100\, {\rm ns}$.  
This indicates that the decoherence time of a single spin  
 is even larger, $T_2\geq T_2^*$.  
However, the mechanisms of spin decoherence for extended and localized  
 electrons are rather different (cf. Ref.~\onlinecite{Pikus})   
 and, therefore, the result of Ref.~\onlinecite{Kikkawa}   
 is not conclusive for electrons in QDs.  
  
Different mechanisms of spin relaxation in QDs have been   
 considered, such as spin-phonon coupling via  
 spin-orbit~\cite{Khaetskii,Khaetskii2} or hyperfine   
 interaction~\cite{Erlingsson}, and   
 spin-nuclear coupling~\cite{BLD,KhaetskiiLossGlazman,MerculovEfrosRosen}.  
The spin-orbit mechanisms yield no spin decay at $B=0$,   
 due to the Kramers degeneracy of the dot states.  
Interestingly, for GaAs QDs, the orbital effect of $B$   
 leads to no spin decay in lowest order in spin-orbit   
 interaction~\cite{Khaetskii,Khaetskii2,Halperin,AleinerFalko}.   
This is due to the special form (linear in momentum) of   
 the spin-orbit coupling in two-dimensions (2D).  
The leading order contribution is, thus, proportional to the Zeeman   
 splitting and leads to long spin-flip times $T_1$ in GaAs QDs
  varying strongly with the $B$-field~\cite{Khaetskii}. 
However, previous theories do not apply to the high values of  $B$
 used in recent experiments~\cite{Hanson}, and thus, no comparison  
 could be made so far.   
As for the nuclear mechanism, the electron spin decay   
 can be suppressed by applying a $B$-field or by polarizing the nuclear   
 system~\cite{BLD,KhaetskiiLossGlazman}.  
  
In this Letter, we show that the spin $T_2$ time, caused by  
 spin-orbit interaction in GaAs QDs, is as large  
 as the spin $T_1$ time.  
We assume low temperatures, $T\ll\hbar\omega_0$,   
 where $\hbar\omega_0$ is the size-quantization energy of the dot,  
 and with no external noise in the applied magnetic field $B$.  
We, thus, argue that the lower bound $T_1\geq 50\,\mu{\rm s}$ 
 established in Ref.~\onlinecite{Hanson} is, in fact, 
 also a lower bound for $T_2$.  
Furthermore, we show that the spin decay can be reduced by  
 a special choice of direction of ${\bf B}$, if there is Rashba   
 spin-orbit interaction. 
 
The Hamiltonian describing the electron in a QD reads  
\begin{eqnarray}\label{H0}  
H&=&H_d+H_Z+H_{SO}+U_{ph}, \\  
\label{Hd}  
H_d&=&\frac{p^2}{2m^*}+U({\bf r}),\\  
\label{HSO}  
H_{SO}&=&\beta(-p_x\sigma_x+p_y\sigma_y)  
+\alpha(p_x\sigma_y-p_y\sigma_x),\\  
H_Z&=&\frac{1}{2}g\mu_B{\bf B}\cdot\mbox{\boldmath $\sigma$},  
\label{HZ}  
\end{eqnarray}  
 where   
 ${\bf p}=-i\hbar {\mbox{\boldmath $ \nabla$}} +(e/c)  
  {\bf A}({\bf r})$ is the electron 2D kinetic momentum,   
 $U({\bf r})$ is the lateral confining potential, with ${\bf r}=(x,y)$,   
 and $\mbox{\boldmath $\sigma$}$ are the Pauli matrices.  
The axes $x$ and $y$ point along the main crystallographic   
 directions in the $(001)$ plane of GaAs.  
The spin-orbit coupling is given by the Hamiltonian (\ref{HSO}),  
 where the term proportional to $\beta$ originates from the  
 bulk Dresselhaus spin-orbit coupling, which is due to  
 the absence of inversion symmetry in the GaAs lattice;  
 the term proportional to $\alpha$ represents the Rashba spin-orbit coupling,  
 which can be present in quantum wells if the confining potential   
 (in our case along the z-axis) is asymmetric.  
The magnetic field   
 ${\bf B}=B(\sin\theta\cos\varphi,\sin\theta\sin\varphi,\cos\theta)$   
 defines the spin quantization axis via the Zeeman term (\ref{HZ}).  
The phonon potential  is given by  
\begin{equation}  
U_{ph}({\bf r})=
\sum_{{\bf q}j}{ F(q_z)e^{i{\bf q}_\parallel{\bf r}}\over 
\sqrt{2\rho_c\omega_{{q}j}/\hbar} } 
(e\beta_{{\bf q}j}-iq\Xi_{{\bf q}j})  
(b_{-{\bf q}j}^\dag+b_{{\bf q}j})\nonumber\\  
\label{Uph}  
\end{equation}  
 where $b_{{\bf q}j}^\dag$ creates an acoustic phonon with wave vector  
 ${\bf q}=({\bf q}_\parallel,q_z)$, branch index $j$, and dispersion   
 $\omega_{{q}j}$; $\rho_c$ is the sample density   
 [volume is set to unity in (\ref{Uph})].  
The factor $F(q_z)$ equals unity for  
 $|q_z|\ll d^{-1}$ and vanishes for $|q_z|\gg d^{-1}$, where $d$ is the   
 characteristic size of the quantum well along the $z$-axis.  
We take into account both piezo-electric ($\beta_{{\bf q}j}$)   
 and deformation potential ($\Xi_{{\bf q}j}$)   
 kinds of electron-phonon interaction.  
For $\beta_{{\bf q}j}$ and $\Xi_{{\bf q}j}$,  
 we refer the reader to Ref.~\onlinecite{GantmakherLevinson}.  
Next, we derive an effective Hamiltonian, which describes the spin 
 dynamics, decoherence and relaxation at low temperatures 
 $T\ll\hbar\omega_0$.

The electron spin couples to phonons due to the spin-orbit interaction   
 (\ref{HSO}).  
For typical GaAs QDs the spin-orbit length  
 $\lambda_{SO}=\hbar/m^*\beta$ is much larger than the electron   
 orbit size $\lambda$.  
In 2D, there is no linear in $\lambda/\lambda_{SO}$ contribution to the   
 spin-phonon coupling at zero Zeeman   
 splitting~\cite{Khaetskii,Khaetskii2,Halperin,AleinerFalko}.  
Here, we consider a finite magnetic field,   
 $m^*\beta^2\ll g\mu_B B\ll\hbar\omega_0$, for which the linear  
 in $\lambda/\lambda_{SO}$ contribution to the  
 spin--phonon coupling dominates the spin decay.  
Using perturbation theory (or Schrieffer-Wolff transformation),   
 we obtain the effective Hamiltonian~\cite{note1}   
\begin{eqnarray}\label{Heff}  
&&H_{\rm eff}=\frac{1}{2}g\mu_B({\bf B}+\delta{\bf B}(t))  
\cdot\mbox{\boldmath $\sigma$},\\  
&&\delta{\bf B}(t)=2{\bf B}\times\mbox{\boldmath $\Omega$}(t),  
\label{dB}  
\end{eqnarray}  
 where $\mbox{\boldmath $\Omega$}(t)$   
 is defined as follows  
\begin{equation}\label{Omega}  
\mbox{\boldmath $\Omega$}(t)=\langle\psi|  
\left[\left(\hat{L}_d^{-1}  
\mbox{\boldmath $\xi$}\right),U_{ph}(t)\right]|\psi\rangle.  
\end{equation}  
Here, $|\psi\rangle$ is the electron orbital wave function,  
 $\hat{L}_d$ is the dot Liouvillean, $\hat{L}_dA=[H_d,A]$.   
The vector $\mbox{\boldmath $\xi$}$ has a simple form   
 in the coordinate frame: $x'=(x+y)/\sqrt{2}$, $y'=-(x-y)/\sqrt{2}$, $z'=z$,  
 namely, $\mbox{\boldmath $\xi$}=(y'/\lambda_-,x'/\lambda_+,0)$,  
 where $1/\lambda_\pm=m^*(\beta\pm\alpha)/\hbar$.  
Eq.~(\ref{dB}) contains one of our main results:  
 {\em in 1st order in spin-orbit interaction, there can be only   
 transverse fluctuations of the effective magnetic field, 
 i.e. $\delta{\bf B}(t)\cdot {\bf B}=0$.}  
This statement holds true for spin coupling to any fluctuations,  
 be it the noise of a gate voltage or coupling to particle-hole excitations   
 in a Fermi sea.  
Next, we consider the decay of the electron spin,   
 ${\bf S}=\mbox{\boldmath $\sigma$}/2$, governed by   
 the Hamiltonian (\ref{Heff}).  
  
The phonons which are emitted or absorbed by the electron leave the dot  
 during a time $\tau_c$, $d/s\lesssim\tau_c\lesssim \lambda/s$,   
 where $s$ is the sound velocity.  
The electron spin decays over a much longer time span in typical structures,  
 and, therefore, undergoes many uncorrelated scattering events.  
In this regime, the dynamics and decay of the spin is governed
 by the Bloch equation~\cite{Slichter}  
\begin{equation}\label{Bloch}  
\langle{\bf\dot{S}}\rangle={\mbox{\boldmath $\omega$}}  
\times\langle{\bf S}\rangle-  
\Gamma\langle{\bf S}\rangle+{\mbox{\boldmath $\Upsilon$}},  
\end{equation}  
 where $\mbox{\boldmath $\omega$}=\omega{\mbox{\boldmath $l$}}$, with  
 $\omega=g\mu_BB/\hbar$ and ${\mbox{\boldmath $l$}}={\bf B}/B$.  
For a generic $\delta{\bf B}(t)$,  we find that in the Born-Markov   
 approximation~\cite{Slichter,LossDiVin2003}
 the tensor $\Gamma_{ij}$  can be written as   
 $\Gamma=\Gamma^r+\Gamma^d$,   
 with   
\begin{eqnarray}  
&&\Gamma^r_{ij}=  
\delta_{ij}\left(\delta_{pq}-l_pl_q\right)  
J_{pq}^{+}(\omega)  
-(\delta_{ip}-l_il_p)J_{pj}^{+}(\omega)\nonumber\\  
&&\hspace{1cm}  
-\delta_{ij}\varepsilon_{kpq}l_kI_{pq}^{-}(\omega)  
+\varepsilon_{ipq}l_pI_{qj}^{-}(\omega),  
\label{Gr}\\  
&&\Gamma^d_{ij}=  
\delta_{ij}l_pl_q  
J_{pq}^{+}(0)-  
l_il_pJ_{pj}^{+}(0),  
\label{Gd}  
\end{eqnarray}  
 where   
 $J^{\pm}_{ij}(w)={\rm Re}\left[J_{ij}(w)\pm J_{ij}(-w)\right]$   
 and $I^{\pm}_{ij}(w)={\rm Im}\left[J_{ij}(w)\pm J_{ij}(-w)\right]$  
 are given by the spectral function   
\begin{equation}\label{Jij}  
J_{ij}(w)=\frac{g^2\mu_B^2}{2\hbar^2}  
\int_{0}^{+\infty}\left\langle\delta B_i(0)\delta B_j(t)\right\rangle  
e^{-iw t}dt.  
\end{equation}  
The inhomogeneous part in Eq.~(\ref{Bloch}) is given by  
\begin{eqnarray}  
2\Upsilon_i&=&  
l_jJ_{ij}^{-}(\omega)-l_iJ_{jj}^{-}(\omega)  
+\varepsilon_{ipq}I_{pq}^{+}(\omega)  
\nonumber\\  
&&  
+\varepsilon_{iqk}l_kl_p  
\left[I_{pq}^{+}(\omega)-I_{pq}^{+}(0)\right],  
\label{inhomogen}  
\end{eqnarray}  
 where $\varepsilon_{ijk}$ is the anti-symmetric tensor 
(with Einstein summation convention) and we have assumed  
 that $\langle\delta{\bf B}(t)\rangle=0$.  
 Eq.~(\ref{Bloch}) describes spin decay in a number of problems, such as  
 electron scattering off impurities in bulk systems,  
 nuclear spin scattering~\cite{Slichter}, etc.  
In our notation,   
 the spin decay comes from the symmetric part of $\Gamma$,  
 whereas the anti-symmetric part leads to a correction to   
 $\mbox{\boldmath $\omega$}$ in Eq.~(\ref{Bloch}).  
The tensor $\Gamma^r$ describes spin decay due to processes of  
 energy relaxation, such as, e.g., emission/absorption of a phonon.  
Therefore, the $T_1$ time is entirely determined by $\Gamma^r$,   
 see below.  
The tensor $\Gamma^d$ can be non-zero only due to elastic scattering of  
 spin, i.e. due to dephasing.  
 $\Gamma^d$ contributes to the decoherence time $T_2$, and 
 so does $\Gamma^r$.  
In many cases, however, the latter contribution is negligible, and  
 $\Gamma^d$ entirely dominates the spin decoherence~\cite{Slichter}.  
This is in strong contrast to what we find here for an electron localized  
 in a QD.  
To illustrate this, we first consider an example when   
 $\Gamma^d$ dominates the spin decoherence and then return to our case.  
A textbook example is  
 $\langle\delta B_i(0)\delta  
 B_j(t)\rangle=\bar{b}^2\delta_{ij}\exp(-|t|/\tau_c)$.  
Choosing 
 $\mbox{\boldmath $l$}=(0,0,1)$,  
 we  obtain from Eqs.~(\ref{Gr}) - 
 (\ref{Jij}) the non-zero elements:   
 $\Gamma_{xx}^r=\Gamma_{yy}^r=\Gamma_{zz}^r/2  
 =\gamma_n^2\bar{b}^2\tau_c/(1+\omega^2\tau_c^2)$,  
 and $\Gamma_{xx}^d=\Gamma_{yy}^d=\gamma_n^2\bar{b}^2\tau_c$, where   
 $\gamma_n=g\mu_B/\hbar$.  
The longitudinal component $\langle S_z\rangle$ decays  
 over the time $T_1=\Gamma_{zz}^{-1}=1/\Gamma_{zz}^r$.  
The transverse components decay over the time  
 $T_2=1/(\Gamma_{xx}^r+\Gamma_{xx}^d)$.  
At $\omega\gg 1/\tau_c$, the contribution of $\Gamma_{xx}^r$ to  
 $T_2$ is negligible, and hence, $T_2\ll T_1$.  
The latter relation has widely been quoted in the literature  
 on  quantum computation.  
In stark contrast to this example, we show now below that 
 there are no {\em intrinsic} dephasing mechanisms for our case,  
 which would justify this relation for the electron spin in GaAs   
 QDs at $T\ll\hbar\omega_0$.    
   
\begin{figure}\vspace{0.5cm}\narrowtext  
{\epsfxsize=7cm  
\centerline{{\epsfbox{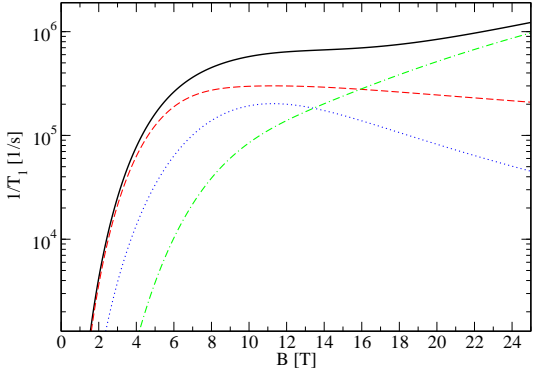}}}}  
\caption{
Solid curve: the relaxation rate $1/T_1$, given in   
Eq.~(\ref{finaltime}), as a function of  
 the $B$-field for a GaAs QD with  
 $\hbar\omega_0=1.1\,{\rm meV}$,   
 $\lambda_{SO}=\hbar/m^*\beta=1\,{\mu}m$,   
 and $\alpha=0$.  
The direction of ${\bf B}$ is in plane, ${\bf B}=(B_x,B_y,0)$.  
The dashed (dotted) curve shows the contribution of the transverse 
 (longitudinal) phonons interacting with the electron via   
 the piezo-electric ($\overline{\beta}_{j\vartheta}$) mechanism.  
The dot-dashed curve shows the contribution of the  
 longitudinal phonons interacting via the deformation potential   
 ($\overline{\Xi}_{j}$) mechanism.  
The angular dependence of $1/T_1$ is given by Eq.~(\ref{ellipse2}),  
 the square of an ellipsoid.  
 }  
\label{relaxrate}  
\end{figure}

We start with calculating the spin decay due to the mechanism (\ref{dB}).    
Here, $\Gamma_{ij}^d$ is identically zero, due to the transverse  
 nature of the fluctuating field $\delta {\bf B}$.  
This can be inferred from Eqs.~(\ref{Gd}) and (\ref{Jij}),  
 noticing that each term in (\ref{Gd})   
 contains $\mbox{\boldmath $l$}\cdot\delta{\bf B}=0$.  
In order to calculate  $\Gamma_{ij}^r$, we first find the  
 main axes of the tensor $J_{ij}(w)$, see Eq.~(\ref{Jij}).  
 $J_{ij}(w)$ is diagonal in the  frame $(X,Y,Z)$, which is obtained  
 from $(x',y',z)$ by a rotation with Euler angles  
 $\varphi'$, $\theta$, and $\chi$. 
Here, the angles $\varphi'=\varphi-\pi/4$ and $\theta$ 
 give ${\bf B}$ in the frame $(x',y',z)$, and $\chi$ depends 
 on the details of the dot confining potential $U({\bf r})$. 
It can be determined from $\langle\delta B_X\delta B_Y(t)\rangle=0$. 
Assuming  $U({\bf r})=U(r)$, we find~\cite{note3}  
\begin{equation}  
\tan2\chi=  
\frac{2\left(\lambda_+^2-\lambda_-^2\right)l_{x'}l_{y'}l_z}
{\lambda_+^2\left(l_{y'}^2-l_z^2l_{x'}^2\right)+  
\lambda_-^2\left(l_{x'}^2-l_z^2l_{y'}^2\right)}.  
\label{chi}  
\end{equation}  
We now consider a harmonic confinement,  
 $U(r)=m^*\omega_0^2r^2/2$, and evaluate   
 $\mbox{\boldmath $\Omega$}(t)$ of Eq.~(\ref{Omega}) for   
 the ground state   
 $\psi({\bf r})=\exp\left(-r^2/2\lambda^2\right)  
 /\lambda\sqrt{\pi}$, where   
 $\lambda^{-2}=\hbar^{-1}\sqrt{(m^*\omega_0)^2+(eB_z/2c)^2}$.  
Using the identity~\cite{note2}  
\begin{equation}\label{y}  
y=\frac{2}{m^*\omega_0^2}\hat{L}_d\left(\frac{\partial}{\partial y}+  
\frac{ieB_z}{\hbar c}x\right),  
\end{equation}  
 we find an expression for $\Omega_{x'}$, which can be obtained from  
 the r.h.s. of Eq.~(\ref{Uph})  
 by replacing   
\begin{equation}\label{subst}  
\exp\left(i{\bf q}_\parallel {\bf r}\right)\to   
\frac{2iq_{y'}}{m^*\omega_0^2\lambda_-}  
\exp\left(-q_\parallel^2\lambda^2/4\right).  
\end{equation}   
Furthermore, an expression for $\Omega_{y'}$ is obtained  
 in the same way, using the substitution (\ref{subst})   
 with a different prefactor, namely with   
 $(q_{y'}/\lambda_-)\to(q_{x'}/\lambda_+)$.  
Finally, for the real part of the correlator $J_{ij}(w)$, we obtain  
\begin{eqnarray}  
\hspace{-1.1cm}&&{\rm Re}J_{XX}(w)=  
\frac{\omega^2w^3(N_w+1)}{(\Lambda_+m^*\omega_0^2)^2}  
\sum_{j}\frac{\pi\hbar^3}{\rho_0s_j^5}\int_0^{\pi/2}d\vartheta\sin^3
\vartheta\nonumber\\  
\hspace{-1.1cm}&&  
\times e^{-(w\lambda\sin\vartheta)^2/2s_j^2}  
\left|F\left(\frac{|w|}{s_j}\cos\vartheta\right)\right|^2  
\left(e^2\overline{\beta}_{j\vartheta}^2+  
\frac{w^2}{s_j^2}\overline{\Xi}_j^2\right),  
\label{Jxx}  
\end{eqnarray}  
 where $N_w=\left(e^{\hbar w/T}-1\right)^{-1}$,  
 and $s_j$ is the sound velocity for branch $j$.   
For GaAs, we use $s_1\approx 4.7\times 10^5\,{\rm cm}/{\rm s}$ and   
 $s_2=s_3\approx 3.37\times 10^5\,{\rm cm}/{\rm s}$.  
Furthermore,  
 $\overline{\Xi}_j=\delta_{j,1}\Xi_0$ with $\Xi_0\approx 7\,{eV}$,  
 and $\overline{\beta}_{1,\vartheta}=  
 3\sqrt{2}\pi h_{14}\kappa^{-1}\sin^2\vartheta\cos\vartheta$,  
 $\overline{\beta}_{2,\vartheta}=  
 \sqrt{2}\pi h_{14}\kappa^{-1}\sin2\vartheta$,  
 $\overline{\beta}_{3,\vartheta}=  
 \sqrt{2}\pi h_{14}\kappa^{-1}(3\cos^2\vartheta-1)\sin\vartheta$,  
 with $h_{14}\approx 0.16\,{\rm C}/{\rm m}^2$ and $\kappa\approx 13$.  
The effective spin-orbit length $\Lambda_{+}$ in Eq.~(\ref{Jxx}) is given by  
\begin{equation}  
\frac{2}{\Lambda_{\pm}^2}  
=\frac{1-l_{x'}^2}{\lambda_-^2}+\frac{1-l_{y'}^2}  
{\lambda_+^2}  
\pm\sqrt{\left(\frac{1-l_{x'}^2}{\lambda_-^2}+\frac{1-l_{y'}^2}  
{\lambda_+^2}\right)^2  
-\frac{4l_z^2}{\lambda_+^2\lambda_-^2}}.  
\label{Lamb}  
\end{equation}  
 ${\rm Re}J_{YY}(w)$ is obtained from Eq.~(\ref{Jxx}) by substituting  
 $\Lambda_+\to\Lambda_-$, and $J_{ZZ}(w)=0$.  
 ${\rm Im}J_{XX}(w)$ and ${\rm Im}J_{YY}(w)$ are irrelevant for our   
 discussion, see further.   
{}From Eq.~(\ref{Gr}),   
 we obtain $\Gamma_{XX}^r=J_{YY}^+(\omega)$,  
 $\Gamma_{YY}^r=J_{XX}^+(\omega)$,  
 $\Gamma_{ZZ}^r=J_{XX}^+(\omega)+J_{YY}^+(\omega)$,  
 $\Gamma_{XY}^r=-I_{YY}^-(\omega)$, $\Gamma_{YX}^r=I_{XX}^-(\omega)$.  
Since   
 $\Gamma_{ij}/\omega\sim \omega\lambda^2/\omega_0\Lambda_\pm^2\ll 1$,  
 we can solve the secular equation iteratively and obtain~\cite{note4}
\begin{eqnarray}  
&&\frac{1}{T_1}:=l_pl_q\Gamma_{pq}=\Gamma_{ZZ},  
\\  
&&\frac{1}{T_2}:=\frac{1}{2}\left(\delta_{pq}-l_pl_q\right)\Gamma_{pq}  
=\frac{1}{2}\left(\Gamma_{XX}+\Gamma_{YY}\right).  
\end{eqnarray}  
Then, the solution of Eq.~(\ref{Bloch}) reads  
 $\langle S_X(t)\rangle=S_\perp e^{-t/T_2}\sin(\omega t+\phi)$,  
 $\langle S_Y(t)\rangle=S_\perp e^{-t/T_2}\cos(\omega t+\phi)$,  
 and   
$\langle S_Z(t)\rangle=S_T+\left(S_Z^0-S_T\right)e^{-t/T_1}$,  
 with the thermodynamic value of spin being   
${\bf S}_T={\mbox{\boldmath $l$}}  
\left({\mbox{\boldmath $l$}}\cdot  
{\mbox{\boldmath $\Upsilon$}}\right)T_1
=-(\mbox{\boldmath $l$}/2)\tanh(\hbar\omega/2k_BT)$,  
 and the initial value   
$\langle{\bf S}(0)\rangle=(S_\perp\sin\phi,S_\perp\cos\phi,S_Z^0)$.  
 For our special situation with purely transverse fluctuations   
($\Gamma^d=0$),   
we obtain   
\begin{equation}  
\frac{1}{T_1}=\frac{2}{T_2}=J_{XX}^+(\omega)+J_{YY}^+(\omega).  
\label{finaltime}  
\end{equation}  
Keeping in mind the setup of Ref.~\onlinecite{Hanson},  
 we plot the relaxation rate $1/T_1$ as a function of $B$  
 for $\theta=\pi/2$ and $\alpha=0$ on Fig.~\ref{relaxrate} (solid curve).  
We find that $1/T_1$ has a plateau in a wide   
 range of magnetic fields (cf. Ref.~\onlinecite{Hanson}),  
 due to a crossover from the piezoelectric-transverse (dashed curve)   
 to the deformation potential (dot-dashed curve)   
 mechanism of electron-phonon interaction.  
For arbitrary $\varphi$, $\theta$, and $\alpha$,   
 we have $1/T_1=f/T_1(\theta=0,\alpha=0)$ with   
\begin{equation}  
f=\frac{1}{\beta^2}  
\left[(\alpha^2+\beta^2)(1+\cos^2\theta)+2\alpha\beta\sin^2\theta  
\sin2\varphi\right]  
\label{ellipse2}  
\end{equation}  
The angular dependence of $1/T_1$ for $\alpha=0$ has been obtained   
 before~\cite{Khaetskii}.  
Note that $\sqrt{f}$ describes an ellipsoid in the frame $(x',y',z)$,   
 i.e. $f={x'}^2+{y'}^2+z^2$, with dimensionless $x',y',z$ obeying  
 $(x'/a)^2+(y'/b)^2+(z/c)^2=1$, where  
 $a=1+\alpha/\beta$, $b=1-\alpha/\beta$, and $c=\sqrt{a^2+b^2}$.  
Note that, if $\alpha=\beta$, then $b=0$, i.e.   
 $1/T_1$ {\em vanishes} if the $B$-field is along $y'$.  
The same is true along the $x'$ axis for $\alpha=-\beta$.  
This special case ($\alpha=\pm\beta$)   
 has been discussed previously for extended electron states  
 in a two-dimensional electron gas (2DEG)~\cite{SchlEugLoss}.  
Note that the Hamiltonian (\ref{H0}) conserves the spin component   
 $\sigma_{y'(x')}$ for $\alpha=\beta$ ($\alpha=-\beta$) and   
 ${\bf B}\parallel y'\,(x')$.  
This spin conservation results in $T_1$ being infinite to all orders  
 in the spin-orbit Hamiltonian (\ref{HSO}).  
At the same time, the decoherence rate $1/T_2$ reduces to  
 the next order contribution of (\ref{HSO}).   
However, as we show below, a single-phonon process is inefficient   
 in inducing dephasing and, therefore, $1/T_2$   
 can be non-zero only in the next order in electron-phonon interaction.  
Next, we note that a long lived spin state also occurs in a different GaAs   
 structure, namely, for a 2DEG grown in the $(110)$ crystallographic   
 direction.   
Then, the normal to the 2DEG plane component of spin is 
 conserved~\cite{Dyak}, provided there is no Rashba coupling.    
  
We discuss now other spin-orbit mechanisms.  
In Eq.~(\ref{HSO}), we did not include  
 the so-called $k^3$-terms of the Dresselhaus spin-orbit 
 coupling~\cite{Dyak}, i.e.   
 $H_{SO}\propto\beta d^2\left(\sigma_x\{k_x,k_y^2\}  
 -\sigma_y\{k_y,k_x^2\}\right)$.  
They are parametrically small ($d^2/\lambda^2\ll 1$) in   
 the 2D limit, compared to the retained ones.   
However, their contribution to the spin decay  
 can be important, if $g\lesssim d^2/\lambda^2$ and   
 $\lambda^2\ll d\lambda_{S0}$, since the orbital effect of the  
 magnetic field contributes here in the first place.  
 Still, for typical GaAs QDs, these mechanisms are negligible.  
  
Additional spin decay mechanisms arise from the {\em direct} spin-phonon  
 interaction~\cite{Khaetskii}.  
The strain field produced by phonons couples to the  
 electron spin via the spin-orbit interaction, resulting  
 in the term   
 $\Delta H'=(V_0/4)\varepsilon_{ijk}\sigma_i\{u_{ij},p_k\}$,  
 where $p_i$ is the bulk kinetic momentum,  
 $u_{ij}$ is the phonon strain tensor,  
 and $V_0=8\times 10^7\,{\rm cm}/{\rm s}$ for GaAs.  
A similar mechanism occurs in a $B$-field,   
 due to $g$-factor fluctuations  
 caused by lattice distortion. This yields  
 $\Delta H''=\tilde{g}\mu_B\sum_{i\neq j}u_{ij}\sigma_iB_j$,  
 where $\tilde{g}\approx 10$ for GaAs.  
The contribution of these mechanisms to the spin-flip rates  
 in QDs has been estimated in Ref.~\onlinecite{Khaetskii}.  
Except for the $\alpha=\pm\beta$ cases discussed above, the  
 direct mechanisms are usually negligible in QDs.   
Here, we find that such spin-phonon couplings do not  
 violate the equality $T_2=2T_1$~\cite{note6}.  
 For this, we note that $\Gamma^d_{ij}=0$ for  
 a generic $\delta B_i=\sum_{\bf q}M_i({\bf q})(b_{-\bf q}^\dag+b_{\bf q})$   
 in Eq.~(\ref{Heff}), if $q\left|M({\bf q})\right|^2\to 0$ at $q\to 0$.  
Obviously, this condition is satisfied for the direct spin-phonon 
 mechanisms, since the strain tensor $u_{ij}$ vanishes at $q=0$.   
The same follows for the Hamiltonian (\ref{H0})  
 with the phonon potential (\ref{Uph}) and an arbitrary $H_{SO}$;  
 the physical explanation being that   
 the potential of long-wave phonons   
 is constant over the dot size and, thus, commutes with $H_{SO}$.  
Finally, we note that, at temperatures $T\sim \hbar\omega_0$,  
 there can be dephasing mechanisms~\cite{Semenov}, which  
 can result in $T_2\ll T_1$.  

In conclusion, we have shown that the decoherence time $T_2$ of an 
 electron spin in a GaAs QD  is as large as the relaxation time $T_1$ 
 for the spin decay based on spin-orbit  mechanisms.
We acknowledge support from the Swiss NSF, NCCR Basel,  
 EU RTN 'Spintronics', US DARPA, ARO, and ONR.


\begin{references}
\bibitem{note1}  
In leading order in $H_{S0}$, we have  
$\tilde{H}\equiv e^SHe^{-S}=H_d+H_Z+U_{ph}+[S,U_{ph}]$, with  
$S$ given by $[H_d+H_Z,S]=H_{SO}$.  
Then, $H_{\rm eff}=\langle\psi|\tilde{H}|\psi\rangle$ up to a spin 
independent  part.
\bibitem{note3}  
We fix the quadrant of $\chi$ by requiring   
that the sign of $\sin 2\chi$ coincides with  
the one of the numerator of Eq.~(\ref{chi}).
\bibitem{note2}  
A similar identity for $x$ is obtained from Eq.~(\ref{y}) by replacing  
$(x,y)$ by $(-y,x)$.
\bibitem{note4}  
The decay rates are defined by the secular equation:  
$\det\left|\left|\Gamma_{ij}-E\delta_{ij}+  
\varepsilon_{ijk}\omega_k\right|\right|=0$, as the real part of $E$.  
In the secular approximation, one iterates the secular equation,  
assuming $\Gamma_{ij}/\omega\ll 1$.
\end{references}
\end{document}